# Size and host-medium effects on topologically protected surface states in bi-anisotropic 3D optical waveguides

**Vasily V. Klimov**[1,2,3,7], **Dmitry V. Guzatov**[4], **Ilya V. Zabkov**[1,2], **Hsun-Chi Chan**[5], and **Guang-Yu Guo**[5,6,8]

[1]*Dukhov Research Institute of Automatics (VNIIA), 22, Sushchevskaya str., 127055, Moscow, Russia*
[2]*P.N. Lebedev Physical Institute, Russian Academy of Sciences, Leninsky Prospekt. 53, 119991, Moscow, Russia*
[3]*National Research Nuclear University MEPhI, Kashirskoe shosse 31, 115409, Moscow Russia*
[4]*Yanka Kupala State University of Grodno, 22 Ozheshko street, 230023, Grodno, Belarus*
[5]*Department of Physics and Center for Theoretical Physics, National Taiwan University, Taipei 10617, Taiwan*
[6]*Physics Division, National Center for Theoretical Sciences, Hsinchu 30013, Taiwan*
[7]*klimov256@gmail.com*
[8]*gyguo@phys.ntu.edu.tw*

**Abstract.** We study the optical properties of bi-anisotropic optical waveguides with nontrivial topological structure in wavevector space, placed in an ordinary dielectric matrix. We derive an exact analytical description of the eigenmodes of the systems with arbitrary parameters that allows us to investigate topologically protected surface states (TPSS) in details. In particular, we find that the TPSS on the waveguides would disappear (1) if their radius is smaller than a critical radius due to the dimensional quantization of azimuthal wavenumber, and also (2) if the permittivity of the host-medium exceeds a critical value. Interestingly, we also find that the TPSS in the waveguides have negative refraction for some geometries. We have found a TPSS phase diagram that will pave the way for development of the topological waveguides for optical interconnects and devices.

## 1. Introduction

Electronic properties of topological phases of matter, including topological insulators, have been under intensive investigation in the past decades [1-4], and this culminated in the Nobel prize in Physics in 2016 being awarded to D. J. Thouless, F. D. M. Haldane and J. M. Kosterlitz "for theoretical discoveries of topological phase transitions and topological phases of matter" [5]. In the meantime, this intensive interest in topological phases of matter has also stimulated widespread studies on complex topology of dispersion relations of photonic crystals and metamaterials, leading to the appearance of topological photonics [6-15, 18-21], a new and vibrant area in nanophotonics and nanooptics. Examples of such photonic materials include bi-anisotropic materials [7], magnetized cold plasma [8, 9] or planar photonics crystals [10-15]. The nontrivial topology in wavevector space can be observed also in ferrite films [16, 17]. Several experimental works of this kind are known both in microwaves [18, 19] and in optics [20, 21].

Among other things, the most interesting effect here is topologically protected surface states (TPSS) whose one-way propagation should be insensitive to the spatial variations of the waveguide surface. This phenomenon sometimes is referred to as bound state in continuum [12], and the corresponding TPSS are often considered as the analogs of conducting surface states in the electronic topological insulators [4].

However, there are some fundamental differences between the surface states on electronic and photonic topological insulators. First, electrons have a mass and a charge and cannot disappear. Second, electrons are subject to Fermi-Dirac statistics, whereas photons to Bose-Einstein statistics. Third and most important difference is that the extension of the electronic wave function is generally much smaller than the system size, and therefore the systems can be considered as infinite half-spaces. On the other hand, the wavelengths of photons in the visible regime, for example, are of the order of 1 μm, and thus can be larger than the transverse size of the optical system, especially optical nano-waveguides. These differences can render the concept of TPSS in photonics invalid in certain circumstances. Furthermore, two-dimensional (2D) propagation of TPSS in a flat interfacial plane or on a flat surface is usually considered in topological photonics. However, one major goal of the topological photonics is the design of long three-dimensional (3D) nanowaveguides with small losses which can be used as optical interconnects [22-23].

To meet this goal in this paper, we present the results of our detailed study of the fundamental properties of TPSS in waveguides of realistic geometries made of photonic metamaterials. Main attention is paid to investigation of how change of waveguide geometry influence properties and even existence of TPSS. In doing so we will consider waveguides of different radius and different cross-sections (see Fig.1). We will also consider how properties of environment influence on existence of TPSS. In particular, we derive an analytical solution for the considered system. Based on this analytical solution, we thoroughly investigate the optical properties especially those which have been overlooked so far, of the TPSS, such as the limited validity of the TPSS concept.

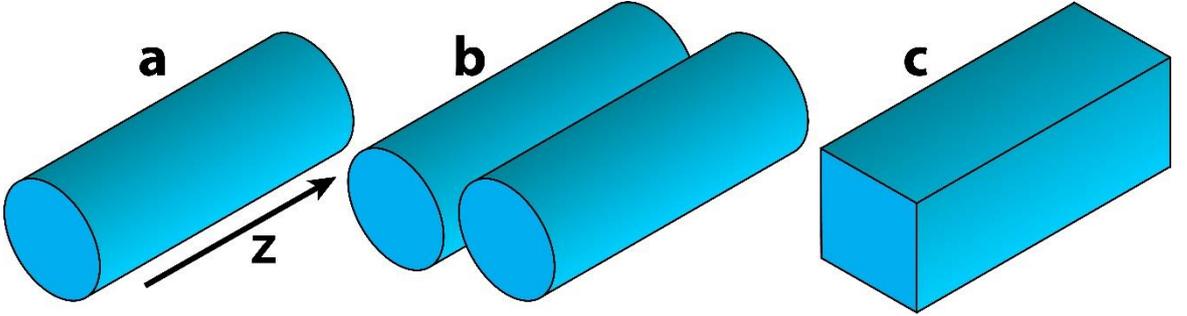

FIG. 1. Geometries under consideration (a) single-wire waveguide; (b) two-wire waveguide and (c) single square waveguide. All waveguides are made of bi-anisotropic material with constitutive equation (1) and have nontrivial topology in wavevector space, while host-medium is an ordinary dielectric with permittivity $\varepsilon_d$.

By ways of example, we focus on the nano-waveguides made of bi-anisotropic material with constitutive relation [24]:

$$\mathbf{D} = \hat{\varepsilon}\mathbf{E} - i\chi\mathbf{H}, \quad \mathbf{B} = \mu\mathbf{H} + i\chi\mathbf{E}, \tag{1}$$

Permittivity tensor $\hat{\varepsilon}$ in (1) has the following form:

$$\hat{\varepsilon} = \begin{pmatrix} \varepsilon_\rho & 0 & 0 \\ 0 & \varepsilon_\rho & 0 \\ 0 & 0 & \varepsilon_z \end{pmatrix}, \tag{2}$$

where $\varepsilon_\rho = \varepsilon_x = \varepsilon_y$ are the components of the permittivity tensor along Cartesian axes *x* and *y*, and $\varepsilon_z$ is the permittivity component along Cartesian axis *z*. Let the medium under consideration has isotropic magnetic permeability $\mu$ and isotropic chirality parameter $\chi$. We also assume that the waveguide is placed in ordinary isotropic dielectric host of permittivity $\varepsilon_d$. When $\varepsilon_\rho > 0$ and $\varepsilon_z < 0$ (hyperbolic case [25]) such medium has nontrivial topology in reciprocal space.

This paper is organized as follows. The derived analytical expression for the dispersion relations for the cylindrical waveguide made of bi-anisotropic material and placed in an ordinary dielectric host is presented in Section 2. Section 3 is devoted to the study of the topology of the surfaces defined by dispersion relations of bulk bi-anisotropic media and associated singularity of wavevectors. Here we have calculated Chern numbers of all bands also. The influence of the waveguide radius $R$ on the TPSS is analyzed in Section 4. In particular, it is shown in this section that there is a critical radius $R_c$ below which the TPSS would disappear due to the dimensional quantization of eigenmodes. The optical properties of TPSS of two-wire system are analyzed in Section 5. The influence of cross-section shape on TPSS is analyzed in Section 6. The effect of permittivity $\varepsilon_d$ of surrounding medium on TPSS is studied in Section 7. Changes in the dispersion relations of the interface due to the $\varepsilon_d$ variation are reported. In particular, it is demonstrated that for a large enough $\varepsilon_d$, TPSS would also disappear. In Section 8 the calculated group velocity of TPSS is reported to show that for certain sets of $R$ and $\varepsilon_d$ values, the group velocity is opposite to the phase velocity, resulting in a negative refractive index. In Section 9 (Conclusions) a generalized phase diagram of TPSS in cylindrical bi-anisotropic waveguides is presented.

## 2. Electromagnetic waves in a cylindrical bi-anisotropic waveguide: An analytic solution.

To find general expressions of fields inside the bi-anisotropic metamaterial waveguide with translational symmetry along the *z* axis, one can integrate over longitudinal wave vector *h*. In this way, general expressions have the following form:

$$\mathbf{E}^{in} = \int_{-\infty}^{\infty} dh \mathbf{E}, \quad \mathbf{H}^{in} = \int_{-\infty}^{\infty} dh \mathbf{H}. \tag{3}$$

In the case of a circular cylindrical waveguides made of a bi-anisotropic metamaterial (1) one can solve wave equations in the cylindrical system of coordinates $0 < \rho < \infty$, $0 \leq \varphi < 2\pi$ and $-\infty < z < \infty$. In this case, we can show that *z*-components of the fields can be written as:

$$E_z = \sum_{n=-\infty}^{\infty} \left( B_n^P J_n(q_P \rho) + B_n^M J_n(q_M \rho) \right) e^{ihz + in\varphi},$$

$$H_z = \sum_{n=-\infty}^{\infty} \left( f B_n^P J_n(q_P \rho) + g B_n^M J_n(q_M \rho) \right) e^{ihz + in\varphi}. \tag{4}$$

where $J_n(x)$ is the Bessel function [26], $B_n^P$ and $B_n^M$ are coefficients which we can find from the boundary conditions, and $q_P, q_M$ are radial wavevectors:

$$q_{\binom{P}{M}} = \sqrt{\left(k_0^2 - \frac{h^2}{\varepsilon_\rho \mu - \chi^2}\right)\left(\frac{(\varepsilon_\rho + \varepsilon_z)\mu}{2} - \chi^2\right) + 2k_0^2(\chi \mp b\mu)},$$

$$a = \frac{\varepsilon_\rho - \varepsilon_z}{4k_0^2 \chi}\left(k_0^2 - \frac{h^2}{\varepsilon_\rho \mu - \chi^2}\right), \quad b = \sqrt{\frac{\varepsilon_z}{\mu} + a\left(a + \frac{2\chi}{\mu}\right)}.$$

(5)

Equations (4) and (5) allows us to find an analytical solution of the problem. The other field components in the matrix form are given as follows:

$$\begin{pmatrix} E_\rho \\ H_\rho \end{pmatrix} = \frac{ih}{A} M_1 \frac{\partial}{\partial \rho}\begin{pmatrix} E_z \\ H_z \end{pmatrix} - \frac{ik_0}{A} M_2 \frac{1}{\rho}\frac{\partial}{\partial \varphi}\begin{pmatrix} -H_z \\ E_z \end{pmatrix},$$

$$\begin{pmatrix} E_\varphi \\ H_\varphi \end{pmatrix} = \frac{ih}{A} M_1 \frac{1}{\rho}\frac{\partial}{\partial \varphi}\begin{pmatrix} E_z \\ H_z \end{pmatrix} + \frac{ik_0}{A} M_2 \frac{\partial}{\partial \rho}\begin{pmatrix} -H_z \\ E_z \end{pmatrix},$$

(6)

where

$$M_1 = \begin{pmatrix} k_0^2(\varepsilon_\rho \mu + \chi^2) - h^2 & 2ik_0^2 \mu \chi \\ -2ik_0^2 \varepsilon_\rho \chi & k_0^2(\varepsilon_\rho \mu + \chi^2) - h^2 \end{pmatrix},$$

$$M_2 = \begin{pmatrix} \mu(k_0^2(\varepsilon_\rho \mu - \chi^2) - h^2) & i\chi(k_0^2(\varepsilon_\rho \mu - \chi^2) + h^2) \\ -i\chi(k_0^2(\varepsilon_\rho \mu - \chi^2) + h^2) & \varepsilon_\rho(k_0^2(\varepsilon_\rho \mu - \chi^2) - h^2) \end{pmatrix},$$

$$A = \left(k_0^2\left(\sqrt{\varepsilon_\rho \mu} - \chi\right)^2 - h^2\right)\left(k_0^2\left(\sqrt{\varepsilon_\rho \mu} + \chi\right)^2 - h^2\right).$$

(7)

General expressions for the fields outside the waveguide in an ordinary dielectric medium of permittivity $\varepsilon_d$ and permeability being equal to unit, can be written in a similar way

$$\mathbf{E}^{out} = \int_{-\infty}^{\infty} dh \mathbf{E}', \quad \mathbf{H}^{out} = \int_{-\infty}^{\infty} dh \mathbf{H}',$$

(8)

$$E_z' = \sum_{n=-\infty}^{\infty} C_n H_n^{(1)}(q\rho) e^{ihz + in\varphi}, \quad H_z' = \sum_{n=-\infty}^{\infty} D_n H_n^{(1)}(q\rho) e^{ihz + in\varphi},$$

$$\begin{pmatrix} E_\rho' \\ H_\rho' \end{pmatrix} = \frac{ih}{q^2}\frac{\partial}{\partial \rho}\begin{pmatrix} E_z' \\ H_z' \end{pmatrix} - \frac{ik_0}{q^2}\frac{1}{\rho}\frac{\partial}{\partial \varphi}\begin{pmatrix} -H_z' \\ \varepsilon_d E_z' \end{pmatrix}, \quad \begin{pmatrix} E_\varphi' \\ H_\varphi' \end{pmatrix} = \frac{ih}{q^2}\frac{1}{\rho}\frac{\partial}{\partial \varphi}\begin{pmatrix} E_z' \\ H_z' \end{pmatrix} + \frac{ik_0}{q^2}\frac{\partial}{\partial \rho}\begin{pmatrix} -H_z' \\ \varepsilon_d E_z' \end{pmatrix}.$$

(9)

where $H_n^{(1)}(x)$ is the Hankel function of the first kind [26], $q = \sqrt{k_0^2 \varepsilon_d - h^2}$, coefficients $C_n$ and $D_n$ can be found from the boundary conditions.

The nontrivial solution of the system of equations at the boundary is possible only when determinant $D$ of the system is equal to zero. This then leads to the dispersion equation:

$$D = \left\{\frac{nh}{R}\left[\sigma(1,-1) + 2ik_0^2\mu\chi f - \frac{A}{q^2}\right] + ik_0q_P\left[\left[\mu f\sigma(-1,-1) - i\chi\sigma(-1,1)\right]\xi_n(q_P R) - \frac{f\psi_n}{q_P}\right]\right\}$$

$$\times\left\{-\frac{nh}{R}\left[g\sigma(1,-1) - 2ik_0^2\varepsilon_\rho\chi - \frac{gA}{q^2}\right] + ik_0q_M\left[\left[\varepsilon_\rho\sigma(-1,-1) + i\chi g\sigma(-1,1)\right]\xi_n(q_M R) - \frac{\varepsilon_d\psi_n}{q_M}\right]\right\}$$

$$-\left\{\frac{nh}{R}\left[\sigma(1,-1) + 2ik_0^2\mu\chi g - \frac{A}{q^2}\right] + ik_0q_M\left[\left[\mu g\sigma(-1,-1) - i\chi\sigma(-1,1)\right]\xi_n(q_M R) - \frac{g\psi_n}{q_M}\right]\right\}$$

$$\times\left\{-\frac{nh}{R}\left[f\sigma(1,-1) - 2ik_0^2\varepsilon_\rho\chi - \frac{fA}{q^2}\right] + ik_0q_P\left[\left[\varepsilon_\rho\sigma(-1,-1) + i\chi f\sigma(-1,1)\right]\xi_n(q_P R) - \frac{\varepsilon_d\psi_n}{q_P}\right]\right\} = 0,$$

(10)

where

$$\xi_n(x) = \frac{J'_n(x)}{J_n(x)}, \quad \psi_n = \frac{A}{q}\frac{H'^{(1)}_n(qR)}{H^{(1)}_n(qR)},$$

$$\sigma(s_1, s_2) = k_0^2(\varepsilon_\rho\mu + s_1\chi^2) + s_2 h^2.$$

(11)

In the special case of $\chi \to 0$ and $\varepsilon_\rho = \varepsilon_z = \varepsilon$, eq. (10) correctly reduces to the dispersion equation of a waveguide made of ordinary material [27]. Equations (4)-(11) are the main results of this work and they allow us to investigate all surface and bulk states in details.

## 3. Properties and singular points of bulk bi-anisotropic material

A band structure $\omega = \omega(k_x, h)$ of a spatially homogeneous bi-anisotropic material can be found with help of the theory of uniaxial bi-anisotropic media [28]. For a material with constitutive relations (1),(2), this equation is

$$\left(\frac{h^2}{(\varepsilon_\rho\mu - \chi^2)} + \frac{k_x^2}{(\varepsilon_z\mu - \chi^2)} - \frac{\omega^2}{c^2}\right)\left(\frac{h^2\varepsilon_\rho}{(\varepsilon_\rho\mu - \chi^2)} + \frac{k_x^2\varepsilon_z}{(\varepsilon_z\mu - \chi^2)} - \varepsilon_\rho\frac{\omega^2}{c^2}\right)$$

$$= \frac{\chi^2}{\mu}\left(\frac{h^2}{(\varepsilon_\rho\mu - \chi^2)} + \frac{k_x^2}{(\varepsilon_z\mu - \chi^2)} + \frac{\omega^2}{c^2}\right)^2,$$

(12)

where $k_x$ and $h$ are the components of the wavevector on the $x$ and $z$ axes correspondingly and we put $k_y = 0$ without loss of generality. This band structure (12) is shown in Fig. 2, and is also plotted for a fixed frequency $\omega = k_0 c$ where $c$ is speed of light (the red curve in Fig. 2) in Fig. 3.

Figure 2 shows clearly that the band structure consists of three sheets (bands), namely, one closed surface centered at $\Gamma$ and the two other open surfaces situated above and below (in $h$ direction) this closed surface. These three bands are separated by two gaps. Interestingly, all three bands are topologically nontrivial because they all have a nonzero Chern numbers, as indicated in

Fig. 2. Calculation of Chern invariants for continuous media is not a trivial task [29, 30]. Here the Berry curvature and the Chern numbers are calculated by using the efficient numerical algorithm reported in [31] and our calculated Chern numbers agree with those reported in [7]. One can see from Fig. 3a that the gaps are caused by adding the isotropic chirality (i.e., $\chi$ becomes nonzero). Furthermore, these gaps are topologically nontrivial. Note that the gaps would open only when all components of chirality tensor become nonzero. For example, the simple uniaxial chirality $\chi_x = \chi_y = 0$, $\chi_z \neq 0$ would not open a gap in wavevector space [32].

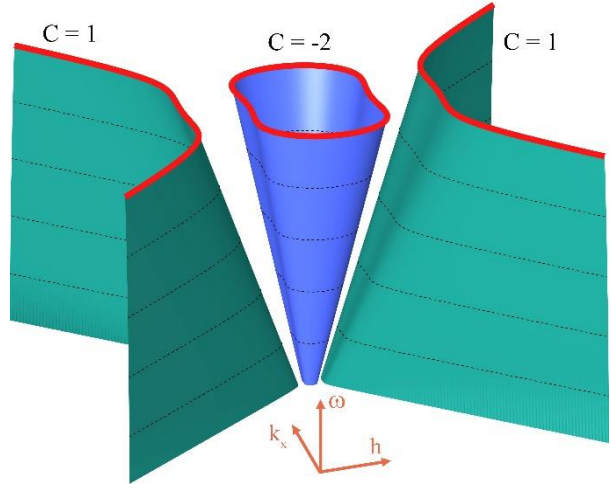

FIG. 2. The band structure $\omega = \omega(k_x, h)$ of a bi-anisotropic medium (12). Here $\varepsilon_\rho = 4$, $\varepsilon_z = -3$, $\mu = 0.5$ and $\chi = 0.5$. The calculated Chern numbers of the surfaces are labelled on the top. The red curve on the top of this diagram is the equifrequency curve plotted again in Fig. 3 below in more details.

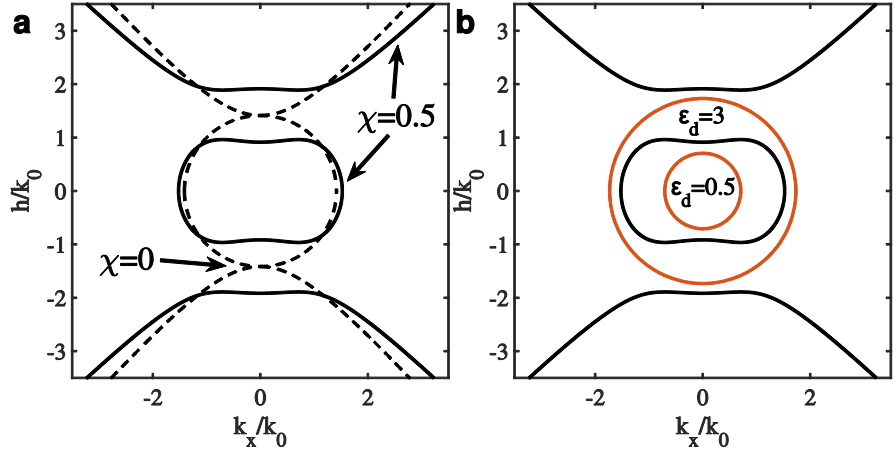

FIG. 3. (a) Equifrequency curves in a bulk bi-anisotropic medium with $\chi = 0$ (dashed lines) and $\chi = 0.5$ (solid lines). (b) Equifrequency lines for an ordinary bulk dielectric with $\varepsilon_d = 0.5$ and $\varepsilon_d = 3$ (red lines). Black lines are the same as in (a). Here $\varepsilon_\rho = 4$, $\varepsilon_z = -3$, $\mu = 0.5$ in both plots.

It is important to note that the equifrequency curve of an ordinary bulk material such as the host dielectric for our waveguide, can lies either inside or outside the equifrequency curves of the bi-anisotropic material or even cross them depending on $\varepsilon_d$. This is demonstrated in Fig. 3b where black lines correspond to the equifrequency curves of a bulk bi-anisotropic material and the red lines to the equifrequency curves of the dielectric with different $\varepsilon_d$ values. As a result, the properties of TPSS can be significantly affected by the host-dielectric and this will be discussed in section 7 below.

Before analyzing modes of bi-anisotropic waveguide [solution of eq. (9)], we should consider all singular points (i.e., the roots) of eq. (9). It is clear from eq. (4) that all singularities of eq. (9) correspond to the zeroes of either wavevector $q_P$ or parameter $b$. They are shown as green and blue lines in Fig. 4. Note that for negative values of $\chi$ wavevector $q_M$ would have zeroes while $q_P$ would not.

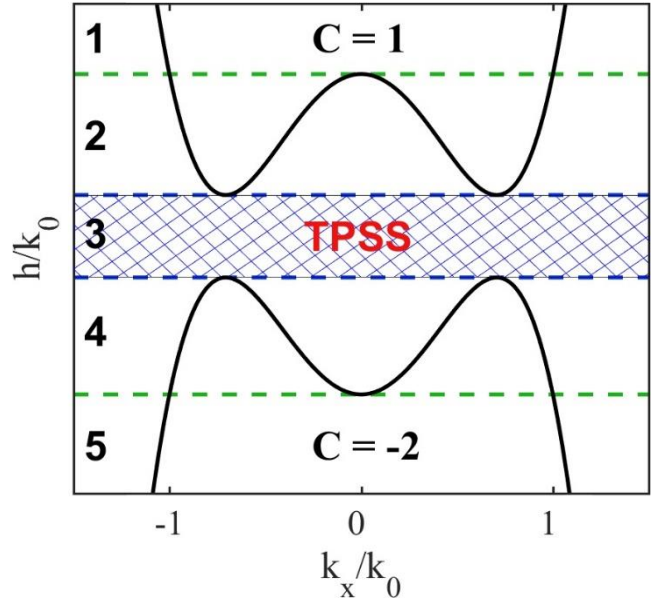

FIG. 4. Singular points in eq. (9) for the eigenmodes of the bi-anisotropic waveguide: $q_P = 0$ (green lines) and $b = 0$ (blue lines). Black lines denote the equifrequency curves of the bulk bi-anisotropic material. Different zones of wavevector $h/k_0$ are marked by numbers on the left. The two band Chern numbers (1 and -2) are shown as well. The hatched area corresponds to possible TPSS.

There are 5 zones which are separated by the dashed green and blue lines. For $h/k_0$ in zones 1 and 5, wavevector $q_P$ is imaginary while $q_M$ is real. For $h/k_0$ in zones 2 and 4, $q_P$ and $q_M$ are real. For $h/k_0$ in zone 3, $q_P$ and $q_M$ have nonzero real and imaginary parts. As a result, bulk modes in the waveguide can in principle appear in zones 1, 2, 4 and 5. Therefore, the only zone where TPSS can exist, is zone 3.

## 4. Influence of the waveguide radius on TPSS properties

It is well known from the theory of usual optical waveguides [33] that for a given frequency, most of the eigen-modes would eventually disappear with decreasing waveguide radius. In this section, let us perform an analogous study on topologically protected surface waves in a circular waveguide made of the bi-anisotropic metamaterial which reveals that TPSS would indeed disappear when its radius becomes sufficiently small.

In order to link the modes in the entire space (Fig. 3) to that of the waveguide, we here introduce the azimuthal wavevector $k_\varphi$:

$$k_\varphi = n/R, \quad (13)$$

where $R$ is the radius of the waveguide. For a planar interface, $k_\varphi \to k_x$. In Fig. 5, solutions of eq. (10) as a function of $h/k_0$ and $k_\varphi/k_0 = n/(k_0 R)$ for different $k_0 R$ values are shown as black crosses. Here we consider the waveguide with the same parameters $\varepsilon_\rho = 4$, $\varepsilon_z = -3$, $\mu = 0.5$ and $\chi = 0.5$ as that considered in [7] for the convenience of comparison. Here bands of the bulk bi-anisotropic material as defined by eq. (12) are shown again as solid black lines.

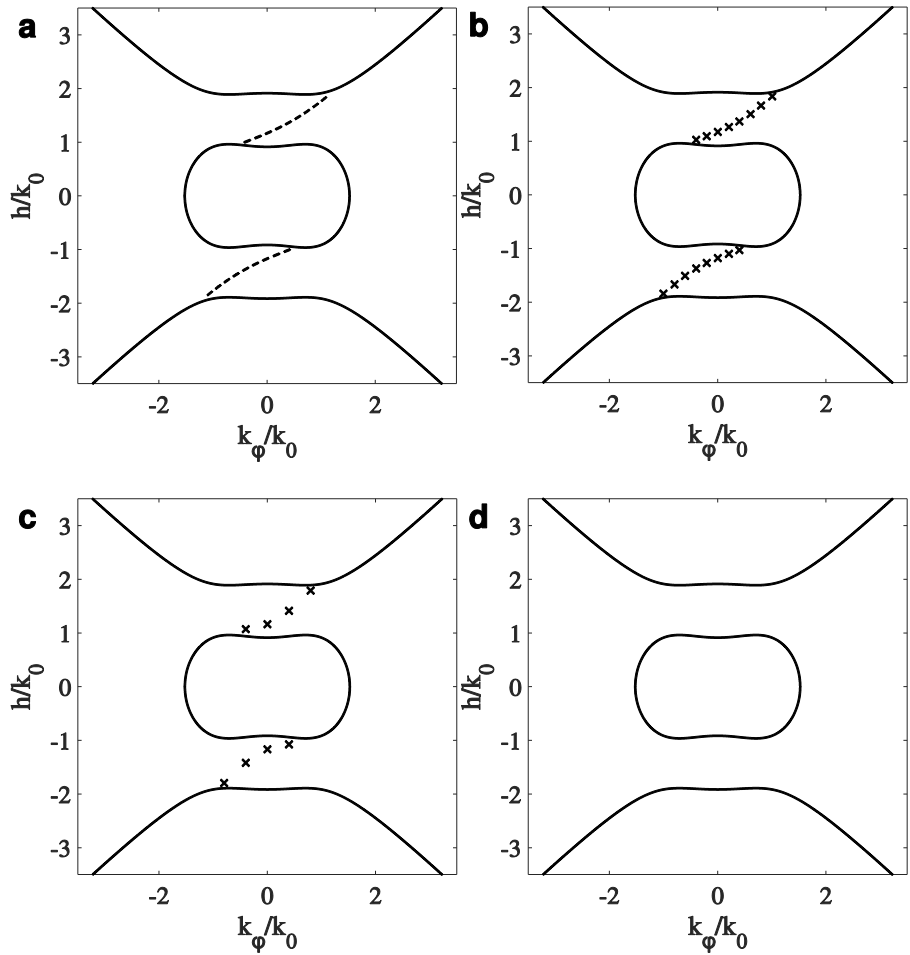

FIG. 5. Eigenvalues of the bi-anisotropic waveguide with different $k_0 R$ values (dashed curves and black crosses) together with equifrequency curves of bulk bi-anisotropic medium [eq. (12)] (solid black lines). (a) Planar interface $k_0 R \to \infty$, (b) $k_0 R = 5$, (c) $k_0 R = 2.5$ and (d) $k_0 R = 0.5$. Note that when $k_0 R = 0.5$ (d) or smaller, there are no TPSS at all.

It is clear that for the bi-anisotropic material filling the half-space, there are topologically protected modes connecting bands of different Chern numbers (12). This corresponds to the waveguide with an infinite radius $k_0 R \to \infty$ (see Fig. 5a). Thus, our results are in good agreement with the result reported in [7]. Similar solutions exist for the waveguide with a finite radius. However, since the cross-section of bi-anisotropic waveguide is finite in this case, the quantization of the modes occurs. The eigenvalues of these modes thus become discrete, as marked by black crosses in Figs. 5b and 5c. Obviously, bigger the radius of the waveguide is, more eigen-modes exist, because more wavelengths can be fit to the circumference. On the other hand, when the radius becomes smaller than a critical radius $R_c$, no TPSS mode would occur (see Fig. 5d for $k_0 R = 0.5$). This is because no eigenmode has a wavelength that can cover the circumference.

Strictly speaking, no TPSS are present in the waveguide with a finite radius. This is because with an infinitesimal change of the waveguide radius $R \to R + \Delta R$, initial longitudinal wavenumber $h$ is no longer a solution of the new dispersion equation and consequently the initial mode disappears through radiation into the environment. This is due to the variations of $\Delta h = h(k_{\varphi 1}) - h(k_{\varphi 2})$ for the neighboring modes. However, in the case of large $k_0 R$ (for $k_0 R \to \infty$), $\Delta h \to 0$ and therefore the effect of topologically protection can be observed in experiment because of the finite bandwidth of real signals.

Selected field distributions of TPSS are shown in Fig. 6. One can see that the azimuthal number $n$ corresponds to the number of maxima of the electric and magnetic fields along the circumference of the waveguide. The field distribution along the waveguide (i.e., the $z$ axis) will be discussed in section 8.

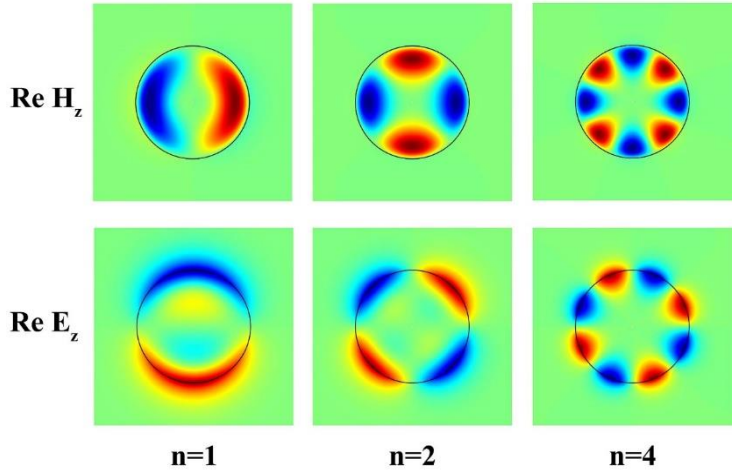

FIG. 6. Distributions of $\mathrm{Re}\,E_z$, $\mathrm{Re}\,H_z$ on the waveguide cross-section for $\chi = 0.5$, $k_0 R = 5$ and $n = 1, 2, 4$. These parameters are marked by the blue crosses in Fig. 8.

Increasing $k_0 R$ would lead to increasing the number of maximums inside the waveguide. However, since the radial wavevectors $q_M$ and $q_P$ are complex, the fields still decay inside the waveguide. Therefore, in order to see how the additional maximums appear, let us plot $\arg E_z$ in Fig. 7 for $k_0 R = 2, 4, 8$, which will show clearly the points for which $E_z = 0$ and how new type of the modes appears.

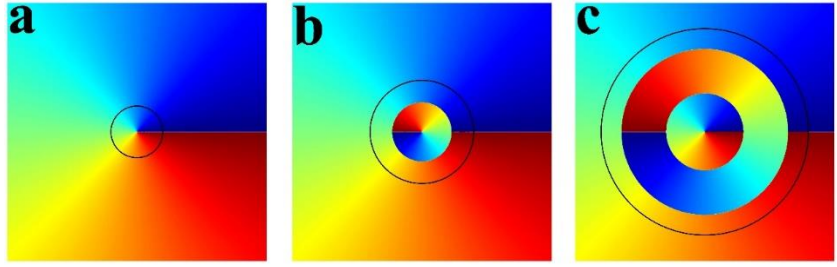

FIG. 7. Distributions of $\arg E_z$ on the waveguide cross-section for $\chi = 0.5$, $n=1$ and $k_0 R = 2, 4, 8$ in (a), (b) and (c), respectively. The parameters used are marked by black crosses in Fig. 8.

Figure 7 shows that increasing $k_0 R$ leads to increasing number of periods of field inside waveguide. Note that changing signs of both $\chi$ and $n$ preserve values of $h$.

Now let us study TPSS dependence on $k_0 R$ in more details. Figure 8 shows the dependence of the modes of the bi-anisotropic waveguide [i.e., solutions of eq. (10)] on $k_0 R$ for $\chi = 0.5$ and the environment with $\varepsilon_d = 1$.

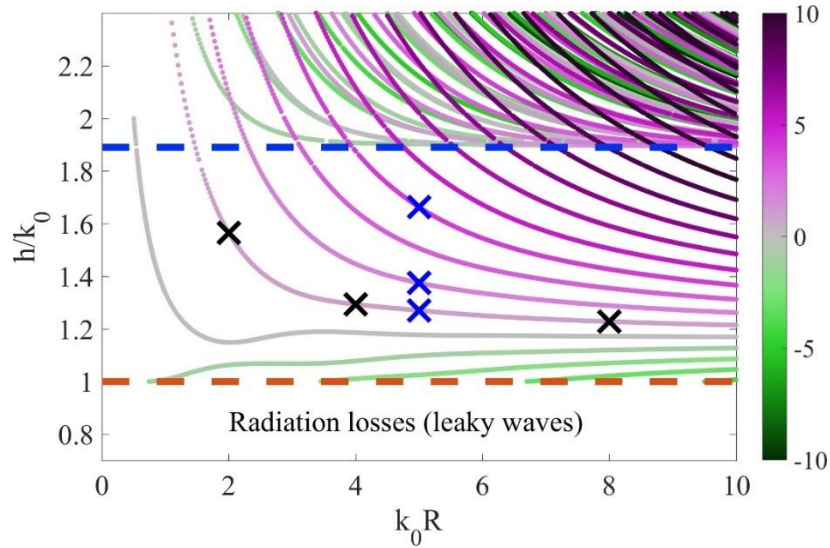

FIG. 8. Dependence of the wavevector $h/k_0$ of the eigenmodes of the bi-anisotropic waveguide on $k_0 R$ for $\varepsilon_d = 1$. Each point denotes a solution of eq. (10). $n$ is shown by color. Crosses correspond to the parameters for which fields are shown in Fig. 6 and Fig. 7.

First of all, one can see from Fig. 8 that all TPSS lie above $h/k_0 = \sqrt{\varepsilon_d}$ line because below this line radiation losses appear. Second, above the blue dashed line, all eigenmodes are bulk ones inside the bi-anisotropic cylinder (see also Fig.4). Thus, the surface modes lie below blue line. One can see that the modes with different signs of azimuthal number $n$ have significantly different dependence of $k_0 R$. For a chosen sign of chirality $\chi = 0.5$, the wavevector $h/k_0$ of the modes with $n < 0$ decrease with decreasing $k_0 R$ (if we do not take into consideration small bends). Moreover, for each $n$ there is a cut-off value for the waveguide radius. In contrast, there is no such cut-off for the modes with $n \geq 0$. However, the longitudinal wavevector $h/k_0$ for these modes increases with decreasing $k_0 R$ and for each mode, there is a critical value of $k_0 R$ below which modes cross the blue line and became bulk high-$k$ hyperbolic modes [25] with large losses, i.e., they disappear as TPSS.

Thus, in this section we have shown that for small enough radius of waveguide TPSS do not exist. It means that if waveguide radius varies along propagation direction and reaches critical values at some distance, the TPSS at this point transform into other modes including radiation ones. In other words, at this point full breakdown of the bulk-edge correspondence [30] will occur.

Until this moment all results were obtained from analytical expressions derived in Section 2. In the following two sections we used numerical simulation in Comsol Multiphysics. To model chirality we adopted method described in [34].

## 5. Topological protected surface states on a two-wire waveguide

As shown in the previous section, TPSS can disappear if the radius of the single-wire waveguide becomes too small. However, from the theory of usual waveguides it is known that change of spatial topology of waveguides can change cutoff effect substantially. In particular, co-axial and two-wire waveguides have no cutoff at all. So, to see how adding another wire to the waveguide would affect the TPSS properties, here we present the results of an investigation of TPSS properties of a two-wire waveguide (see Fig.1b).

Similar to Fig. 8, in Fig. 9 we plot the $h/k_0$ as a function of $k_0 R$ for the two-wire waveguide made of bi-anisotropic materials embedded in the air ($\varepsilon_d = 1$). As usual, modes of two-wire waveguide are symmetric and anti-symmetric superpositions of modes of single waveguide. Consequently, the two-wire system has more TPSS modes for $k_0 R < 2$ [e.g., 5 modes in Fig. 9, 6th mode is leaky] than the single-wire system (only three modes for $k_0 R < 2$ in Fig. 8). From Fig. 9 one can observe that the curves generally behave like the curves in Fig. 8 (which are shown by dashed black lines). In particular, the cutoff radius $k_0 R_c$ still exists in the two-wire system, although it decreases slightly to $k_0 R_c \approx 0.4$. The existence of the cutoff radius indicates that the limitation of the single-wire system remains unchanged even when another wire is added to the single-wire system.

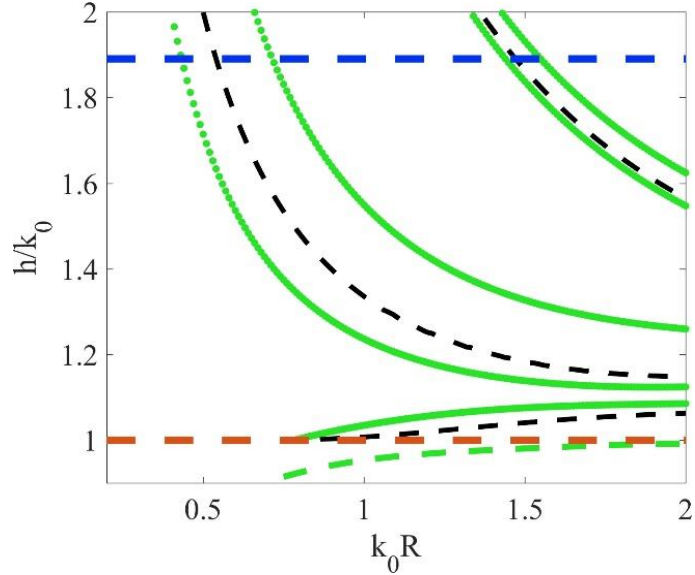

FIG. 9. Wavevector $h/k_0$ as a function of $k_0 R$ for the two-wire system (green curves). The black dashed lines correspond to single-wire case (see Fig.8). The distance between the centers of two cylinders is $2.2R$. Green dashed line symbolized leaky character of mode which appears due to splitting of $n=-1$ mode of one cylinder.

Thus, the results of this section show that two-wire waveguides made of bi-anisotropic metamaterials with nontrivial topology of reciprocal space have almost the same cutoff effects for TPSS as single-wire waveguide.

## 6. TPSS in the waveguide with a superelliptical cross-section

In this section, we will investigate the influence of the shape of the cylinder cross-section on the dimensional quantization. In fact, most of the works on topological photonics investigates this effect. Let us consider cylinders with the cross-section in a superellipse shape [35] which is defined by equation

$$|x|^N + |y|^N = R_c^N. \tag{14}$$

Eigenmodes of such waveguides with different $n$ values and a fixed perimeter $P$ are displayed in Fig. 10 for $k_0 R = 5$. The fixed perimeter condition is used to keep $k_\varphi P = 2\pi n$ where $n$ is an integer. The value of $R_c$ for each $N$ in (14) is chosen so that perimeter $P$ remains the same as for a circle of $R = R_c$. So in all cases $P = 2\pi R$. Perimeter of a superellipse can be found from the following formula [35]):

$$P = 4\sum_{r=0}^{\infty}\sum_{s=0}^{r}\binom{1/2}{r-s}\left(\frac{2(N-1)(r-s)}{N}\right)_s \frac{1}{s!}\frac{2R_c}{2(N-1)(r-s)+Ns+1} \times 2^{-\frac{2(N-1)(r-s)+Ns+1}{N}} \tag{15}$$

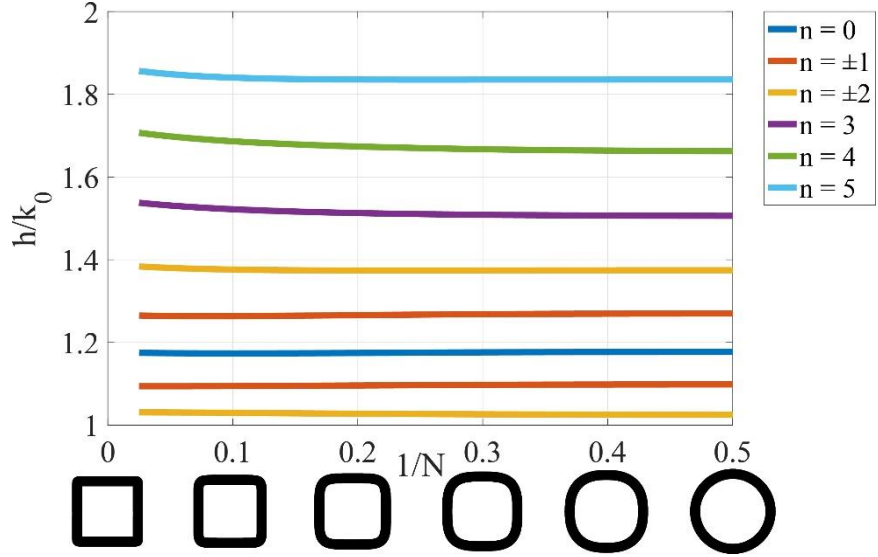

FIG. 10. Longitudinal wavevectors of TPSS modes of a waveguide with a superelliptical cross-section with a fixed perimeter [eq. (14)] for $k_0 R = 5$ as a function of its shape. Bottom row shows the evolution of the cross-section shape from a square to a circle.

One can see from Fig. 10 that for large value $k_0 R \gg 1$ the shape has no significant effect on TPSS and the main parameter is the perimeter of the cylinder. This result agrees with results of most of the works on topological photonics.

## 7. Effect of the environment permittivity on TPSS

Now let us consider how the permittivity $\varepsilon_d$ of the host matrix influences TPSS. Mode structures for different $\varepsilon_d$ values are shown in Fig. 11 for $k_0 R = 20$ and $\chi = 0.5$.

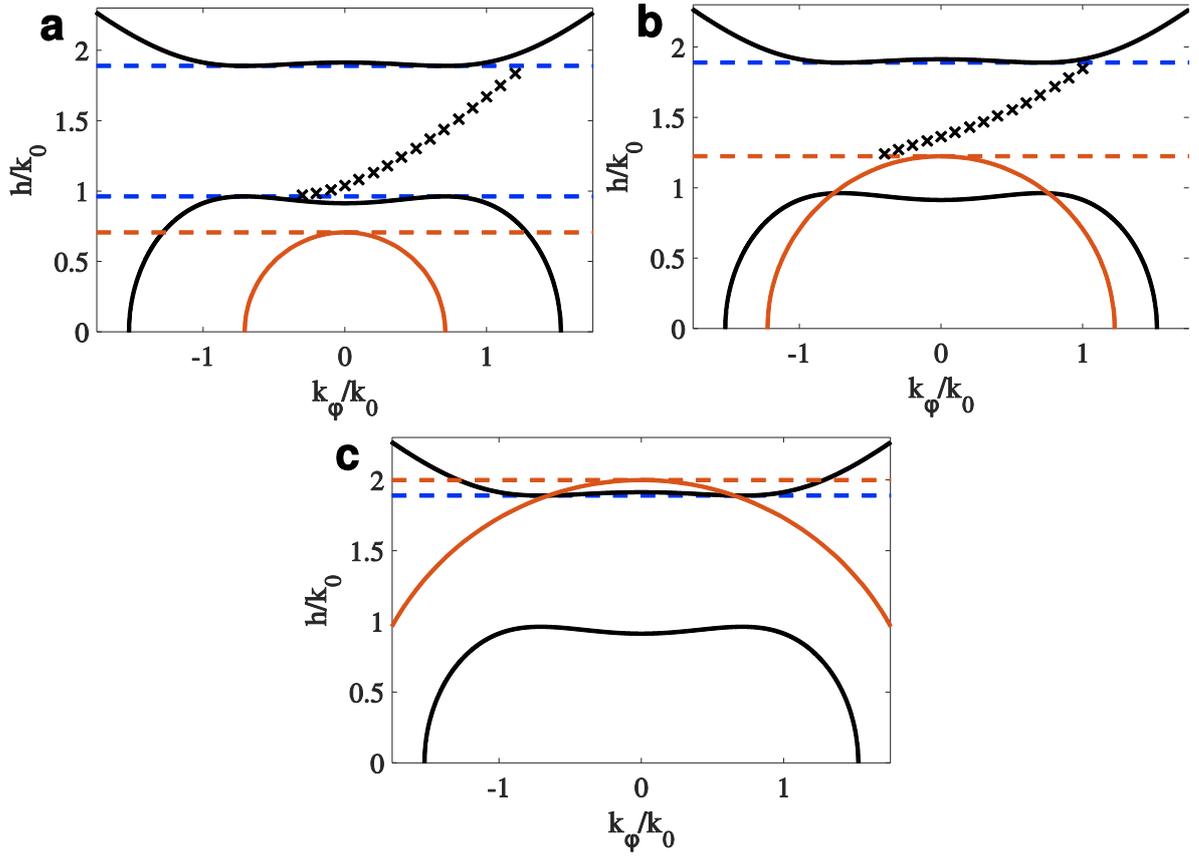

FIG. 11. Eigenvalues [solutions of eq. (10)] of the bi-anisotropic waveguide with $k_0 R = 20$ and with different values of $\varepsilon_d$ (crosses). (a) $\varepsilon_d = 0.5$, (b) $\varepsilon_d = 1.5$, and (c) $\varepsilon_d = 4$. Red semicircles represent the equifrequency surface of the environment.

Analysis of Fig. 11 reveals a significant difference between TPSS and edge states in topological insulators. While the edge states are determined exclusively by the topology in the wavevector space of the bulk topological insulator, TPSS depend significantly on the permittivity of the host medium with trivial topology of isofrequencies surfaces. In the case of

$$\varepsilon_d < \left(\sqrt{\varepsilon_\rho \mu} - |\chi|\right)^2, \tag{16}$$

TPSS arc connects regions of equifrequency surfaces of bulk bi-anisotropic material with different Chern numbers (Fig. 11a, $\varepsilon_d = 0.5$). This situation is similar to topological insulators. In the case of

$$\varepsilon_d > \left(\sqrt{\varepsilon_\rho \mu} - |\chi|\right)^2, \tag{17}$$

TPSS arc starts from the equifrequency surface of the trivial host medium and ends at the equifrequency surface of the bi-anisotropic material with Chern index $C = 1$ (Fig. 11b, $\varepsilon_d = 1.5$). Strictly speaking, in both (a) and (b) cases, TPSS arcs are limited from top by blue line $b = 0$ (see also Fig. 4). Above this line, eigenmodes of the waveguide become the bulk high-$k$ hyperbolic ones. Finally, when equifrequency surface of the host medium crosses the top branch of the bi-anisotropic material, i.e., for $\varepsilon_d > \left(\sqrt{\varepsilon_\rho \mu} + |\chi|\right)^2$, there is no TPSS at all (Fig. 11c, $\varepsilon_d = 4$). It is very important to note that all these conditions are independent of the geometry of the waveguide.

## 8. Negative refraction of TPSS

An important property of TPSS of the waveguide is one-way propagation along the $\varphi$ direction. Usually such propagation is related to the phase velocity. However, from the physical point of view, the group velocity is more important because it is related to the energy flow. Thus, in this section we present the results of our study on how the directions of phase and group velocities of TPSS depend on the parameters of the bi-anisotropic waveguide.

The direction of phase velocity coincides with the direction of wavevector $\mathbf{k}$ which is orthogonal to the phase front of the electromagnetic field. For dependence

$$E_z \sim \exp\left(i\left(hz + n\varphi - \omega t\right)\right) = \exp\left(i\left(hz + k_\varphi l - \omega t\right)\right), \qquad (18)$$

where $l = \varphi R$ is the distance along circumference, one can see that $\mathbf{k} = \left\{0, n/R, h\right\}$. To characterize the direction of group velocity, one should use Poynting vector $\mathbf{S} = \frac{c}{4\pi}[\mathbf{E} \times \mathbf{H}]$ with coordinates $\mathbf{S} = \left\{S_r, S_\varphi, S_z\right\}$ in the cylindrical coordinate system. It can be shown that for TPSS, $S_r = 0$, $S_z$ and $S_\varphi$ do not depend on $\varphi$ coordinate. Streamlines of Poynting vector and phase velocity are shown in Fig. 12 for $k_0 R = 5$, $k_0 R = 1.5$ and $n = -1, 0, 1$.

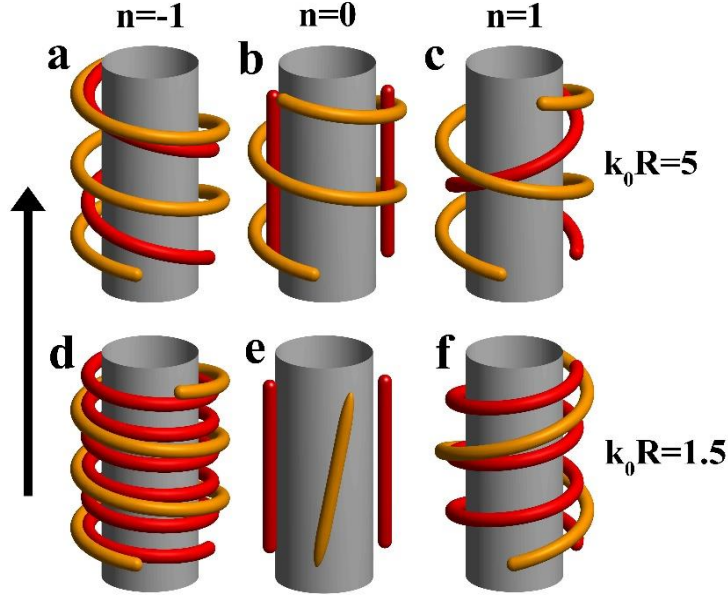

FIG. 12. Streamlines of Poynting vector (orange lines) and phase velocity (red lines) of TPSS with $n = -1$ (a, d), $n = 0$ (b, e), $n = 1$ (c, f) for $k_0 R = 5$ (a, b, c), $k_0 R = 1.5$ (d, e, f) and $\chi = 0.5$. The field distribution and Poynting vector are calculated at $\sqrt{x^2 + y^2} = 1.01R$. In all cases the $z$ component of the velocities is in the positive direction of the $z$-axis.

Figure 12 shows that streamlines of phase velocity have simple behavior, i.e., they change their direction (spirality) together with azimuthal number $n$. On the contrary, directions of group velocity exhibit a more complicated behavior. From Fig. 12 one can see that for $k_0 R = 5$, the stream line of group velocity varies only slightly upon the variation of azimuthal number $n$ from -1 to +1, while for $k_0 R = 1.5$ it varies significantly and even changes its sign. For $k_0 R = 5$ and $n = 1$, group and phase velocities have different signs of $\varphi$ component, while for $k_0 R = 1.5$ and $n = \pm 1$, they have the same sign. To study this effect in more details, we plot in Fig. 13 group velocity (Poynting vector) directions of all TPSS shown in Fig. 8. To do this, we introduce the angle of group velocity spiral $\theta = \arctan(S_\varphi / S_z)$. Positive/negative values of $\theta$ correspond to clockwise/counterclockwise twist of the spiral of Poynting vector (Fig. 12d and Fig. 12f). The values of $\theta$ is also shown in Fig. 13a for TPSS by pseudo-colors. It can be seen that for the modes with $n \geq 0$, angle $\theta$ changes its sign when $k_0 R$ decreases. This change of sign could happen in the TPSS region (zone 3 on Fig. 4). For the chosen optical parameters of the waveguide, it corresponds to the modes with $n = 0, 1, 2, 3, 4$. Therefore, there are values of $k_0 R$ for which one part of the modes would have $\theta < 0$ and the other part $\theta > 0$. An example of such behavior is shown in Fig. 14 for $k_0 R = 5$.

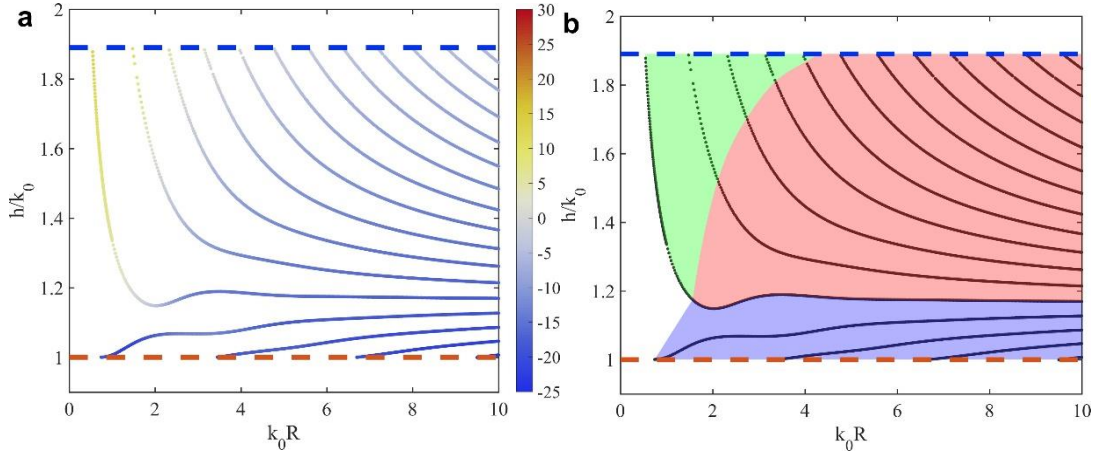

FIG. 13. (a) Direction of group velocity $\theta$, shown by color, for the eigenmodes of the bi-anisotropic waveguide. (b) Three regions of different behavior of phase and group velocities of the eigenmodes. Only TPSS modes are shown.

As it was mentions above sign of $\varphi$ component of phase velocity $\mathbf{v}_{ph}$ is determined by the sign of $n$ (see (18)). On the other hand, $\varphi$ component of group velocity $\mathbf{v}_g$ coincides with $\mathbf{S}_\varphi$ and therefore its sign is determined by the sign of $\theta$. If one consider relative direction of $\varphi$ component of phase and group velocities (which define twist of the spiral) three different regions can be defined (see Fig. 13b). First, for $n < 0$ we have $v_{ph,\varphi} < 0$ and $v_{g,\varphi} < 0$ (region 1 – blue color in Fig. 13b). Second for $n > 0$ and $\theta < 0$ we have $v_{ph,\varphi} > 0$ and $v_{g,\varphi} < 0$ (region 2 – red color in Fig. 13b). Third for $n > 0$ and $\theta > 0$ we have $v_{ph,\varphi} > 0$ and $v_{g,\varphi} > 0$ (region 3 – green color in Fig. 13b). Thus, in the region 2 we have region with negative refractive index along $\varphi$ direction.

TPSS for specific radius $k_0 R = 5$ are shown in Fig. 14. Color of the crosses shows particular region modes belong to. Modes with $n = 0$ is shown by black color since they lie on the border of two regions.

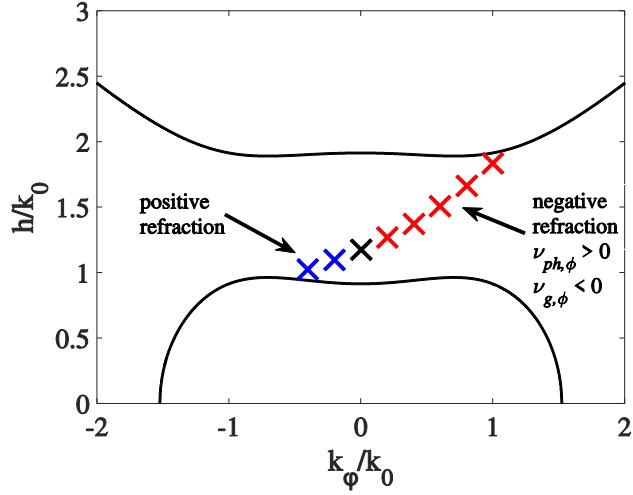

FIG. 14. Eigenvalues (crosses) of the bi-anisotropic anisotropic waveguides for $k_0 R = 5$. Colors of the crosses indicate the regions in Fig. 13b to which the modes belong. Solid lines show bulk equifrequency surfaces.

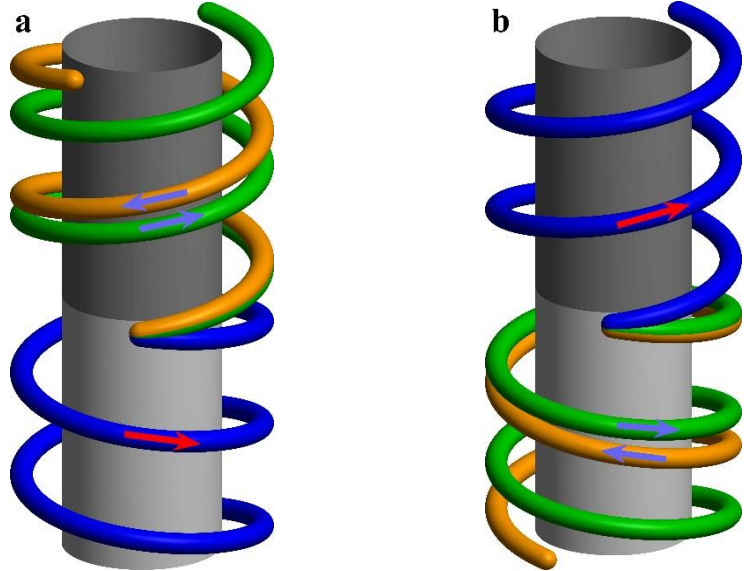

FIG. 15. Light reflection and transmission in TPSS along the interface between two waveguides with different $\chi$ (bottom and top cylinders have $\chi = 0.5$ and $\chi = -0.5$, respectively). Incoming wave is shown by orange color, entering from top in (a) but from bottom (b), while reflected and transmitted waves are shown by green and blue color, respectively.

Existence of negative refraction for TPSS will results in many effects [36-42] and first of all in negative refraction itself (see Fig. 15). From Fig. 15 one can see that negative refraction takes place indeed. In the case of excitation of TPSS with point source one can expect superlens effect [36-42] and so on.

## 9. Conclusions

We have investigated the optical properties of bi-anisotropic waveguides with nontrivial topological structure of reciprocal space surrounded by an ordinary dielectric matrix. In particular, we have derived the exact analytical solution of eigenmodes for the system in the entire parameter space. Our results reveal that the concept of topologically protected photonic surface states (TPSS) has only a limited region of applicability in the parameter space. For example, outside this applicable region such as too small radius of the waveguide, TPSS disappear due to the dimensional quantization of the wavevector. Moreover, permittivity $\varepsilon_d$ of host dielectric matrix is also found to have significant impact on the structure of TPSS, and TPSS could even disappear when $\varepsilon_d$ becomes sufficiently large. This effect is definitely absent in the case of topological insulators, where existences of edge state depends only on topological structure of band zones rather than on topology of environment. The critical value of the waveguide radius below which TPSS vanish, is also shown to depend on the $\varepsilon_d$ value. Based on the analytic solution, we have constructed a TPSS phase diagram in space $\varepsilon_d$ and $1/k_0 R$, as shown in Fig. 16.

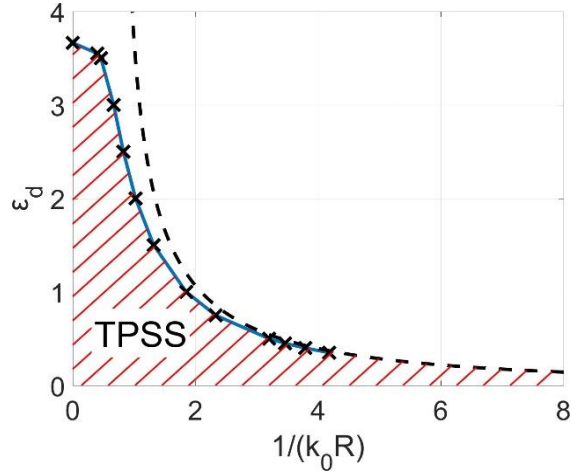

FIG. 16. Phase diagram of TPSS in cylindrical bi-anisotropic waveguides in coordinates $\varepsilon_d$ and $1/k_0 R$. Crosses represent the numerical solutions, while the dashed line denotes the asymptotic analytic solution of (10).

We also have studied phase and group velocities of TPSS, and discover that in certain parameter regions, phase and group velocities of TPSS have opposite signs, i.e., TPSS have negative refractive index. Our interesting findings would certainly be important for designing optical interconnects [22, 23] based on waveguides with nontrivial topological structure of reciprocal space.

## Acknowledgments

The authors are grateful to the Russian Foundation for Basic Research (Grant 15-52-52006 and 18-02-00315) as well as the Ministry of Science and Technology of Taiwan (Grant MOST 104-2923-M-002-004-MY3) and the National Center for Theoretical Sciences of Taiwan for financial support of this work. The numerical simulations of Section 6 were supported by the Russian

Science Foundation (Project No. 18-12-00468). VVK acknowledges support from the MEPhI Academic Excellence Project (Contract No. 02.a03.21.0005).